\documentclass[conference]{IEEEtran}
\IEEEoverridecommandlockouts
\usepackage{cite}
\usepackage{amsmath,amssymb,amsfonts}
\usepackage{graphicx}
\usepackage{textcomp}
\usepackage[table]{xcolor}
\usepackage{bm}
\usepackage{graphicx}
\usepackage{graphics}
\usepackage{algorithm}
\usepackage{algorithmic}
\usepackage{epstopdf}
\usepackage{caption}
\usepackage{subfigure}
\usepackage{amsmath}
\usepackage{amstext}
\usepackage{amsthm}
\usepackage{amssymb}
\usepackage{framed}
\usepackage{multirow}
\usepackage{balance}
\usepackage{mathrsfs}
\usepackage{color}
\usepackage{booktabs} % For formal tables
\graphicspath{{./figures/}}
\newcommand{\argmin}{\arg\!\min}

\newcommand{\RN}[1]{\textup{\uppercase\expandafter{\romannumeral#1}}}
\newcommand{\mat}[1]{{\bf #1}}   % matrix: bold
\newtheorem{problem}{Problem}

\def\BibTeX{{\rm B\kern-.05em{\sc i\kern-.025em b}\kern-.08em
    T\kern-.1667em\lower.7ex\hbox{E}\kern-.125emX}}
\begin{document}

\title{Multi-Level Network Embedding with Boosted Low-Rank Matrix Approximation}

\author{
\IEEEauthorblockN{
Jundong Li,
Liang Wu and
Huan Liu
}
\IEEEauthorblockA{
Computer Science and Engineering, Arizona State University, USA\\
Email: \{jundongl,wuliang,huan.liu\}@asu.edu}}

\maketitle

\begin{abstract}
As opposed to manual feature engineering which is tedious and difficult to scale, network representation learning has attracted a surge of research interests as it automates the process of feature learning on graphs. The learned low-dimensional node vector representation is generalizable and eases the knowledge discovery process on graphs by enabling various off-the-shelf machine learning tools to be directly applied. Recent research has shown that the past decade of network embedding approaches either explicitly factorize a carefully designed matrix to obtain the low-dimensional node vector representation or are closely related to implicit matrix factorization, with the fundamental assumption that the factorized node connectivity matrix is low-rank. Nonetheless, the global low-rank assumption does not necessarily hold especially when the factorized matrix encodes complex node interactions, and the resultant single low-rank embedding matrix is insufficient to capture all the observed connectivity patterns. In this regard, we propose a novel multi-level network embedding framework BoostNE, which can learn multiple network embedding representations of different granularity from coarse to fine without imposing the prevalent global low-rank assumption. The proposed BoostNE method is also in line with the successful gradient boosting method in ensemble learning as multiple weak embeddings lead to a stronger and more effective one. We assess the effectiveness of the proposed BoostNE framework by comparing it with existing state-of-the-art network embedding methods on various datasets, and the experimental results corroborate the superiority of the proposed BoostNE network embedding framework.
\end{abstract}


\section{Introduction}
Learning meaningful and discriminative representations of nodes in a network is essential for various network analytical tasks as it avoids the laborious manual feature engineering process. Additionally, as the node embedding representations are often learned in a task-agnostic fashion, they are generalizable to a number of downstream learning tasks such as node classification~\cite{perozzi2014deepwalk}, community detection~\cite{wang2017community}, link prediction~\cite{grover2016node2vec}, and visualization~\cite{tang2016visualizing}. On top of that, it also has broader impacts in advancing many real-world applications, ranging from recommendation~\cite{wang2018path}, polypharmacy side effects prediction~\cite{zitnik2018modeling} to name disambiguation~\cite{zhang2017name}. The basic idea of network embedding is to represent each node by a low-dimensional vector in which the relativity information among nodes in the original network is maximally transcribed.

A vast majority of existing network embedding methods can be broadly divided into two different yet highly correlated categories. Firstly, early network embedding methods are largely based on the matrix factorization approaches. In particular, methods in this family represent the topological structure among nodes as a deterministic connectivity matrix and leverage low-rank matrix approximation strategies to obtain the node vector representations with the assumption that high-quality node embeddings are encoded in a small portion of latent factors. Various methods differ in the way on how the connectivity matrix is built, typical examples include modularity matrix~\cite{tang2009relational}, Laplacian matrix~\cite{belkin2002laplacian}, $k$-step transition matrix~\cite{cao2015grarep} and higher-order adjacency matrix~\cite{yang2017fast}. Secondly, the recent advances of network embedding research are largely influenced by the skip-gram model~\cite{mikolov2013distributed} in natural language processing. The key idea behind these algorithms is that nodes tend to have similar embedding representations if they co-occur frequently on short random walks over the network~\cite{perozzi2014deepwalk,grover2016node2vec} or are directly connected with each other within certain contexts~\cite{tang2015line,tang2015pte}. As these methods often employ a flexible way to measure the node similarity, they have shown to achieve superior learning performance in many scenarios~\cite{hamilton2017representation,cui2017survey}. Even though the two families of network embedding methods are distinct in nature, recent studies~\cite{qiu2018network} found their inherent correlations and then cast the skip-gram inspired network embedding methods in the matrix factorization framework.

To this end, in this work, we investigate the network embedding problem within the framework of matrix factorization due to its broad generalizability. As mentioned previously, after the closed-form connectivity matrix for each network embedding model is derived, the principled way to permit node embedding representation is to perform low-rank matrix approximation, such as with eigendecomposition, singular value decomposition (SVD) and nonnegative matrix factorization (NMF). Here, the fundamental assumption is that the closed-form connectivity matrix is low-rank (a.k.a. \emph{global low-rank}), and the factorized matrix (which is also low-rank) is sufficient to provide a ``one-size fits all" representation to encode the connectivity information among nodes. However, the widely perceived assumption is untenable in practice and may further hinder us learn effective node embedding representations due to the following reasons. On one hand, the connectivity matrix in real-world scenarios is often not sparse and encodes diverse connectivity patterns among nodes, making the low-rank assumption improper~\cite{lee2013local,lee2014local}. Hence, a global low-rank factorization cannot guarantee a good approximation of the closed-form connectivity matrix. On the other hand, simply relying on the global low-rank property for a single representation for nodes cannot fully explain how the nodes are connected with each other in the network. In a nutshell, it is desired to learn the embedding in a forward stagewise fashion to gradually shift the focus to the unexplained node connectivity behaviors as stages progress.

To address the above problems, we propose a multi-level network embedding framework BoostNE to obtain multiple granularity views (from coarse to fine) of the network over the full spectrum. The proposed multi-level network embedding framework is motivated by the gradient boosting framework~\cite{friedman2001greedy} in ensemble learning, which learns multiple weak learners sequentially and then combines them together to make the final prediction. The main contributions of this paper are summarized as follows:
\begin{itemize}
\item \textbf{Formulation} We systematically examine the fundamental limitations of existing network embedding approaches, especially the methods that directly leverage or can be reduced to the framework of matrix factorization. To alleviate the limitations, we propose to study a novel problem of multi-level network embedding by learning multiple node representations of different granularity.
\item \textbf{Algorithm} We propose a novel network embedding framework BoostNE to identify multiple embedding representations from coarse to fine in a forward stagewise manner. The key idea is to leverage the principle of gradient boosting to successively factorize the residual of the connectivity matrix that is not well explained from the previous stage. As we do not impose any assumption that the original connectivity matrix is global low-rank, the developed multi-level method yields better approximations and further leads to more effective embedding representation.
\item \textbf{Evaluation} We conduct experiments on multiple real-world networks from various domains to compare the proposed BoostNE with existing state-of-the-arts. The results demonstrate the effectiveness of the proposed multi-level network embedding framework BoostNE. Further studies are presented to understand why the proposed BoostNE work and how the number of levels impacts the final embedding representation.
\end{itemize}

The rest of this paper is structured as follows. The problem statement of multi-level network embedding is introduced in Section II. In Section III, we propose a novel framework BoostNE that is able to learn the multi-level network representations from coarse to fine with boosted low-rank matrix approximation. Experimental evaluations on real-world datasets are presented in Section IV with discussions. In Section V, we briefly review related work. The conclusions and future work are presented in Section VI.

\section{Problem Definition}
We first summarize the notations used in this paper. We denote matrices using bold uppercase characters (e.g., $\mat{X}$), vectors using bold lowercase characters (e.g., $\mat{x}$), scalars using normal lowercase characters (e.g., $x$). The $i$-th element of vector $\mat{x}$ is denoted by $x_{i}$. The $i$-th row, the $j$-th column and the $(i,j)$-th entry of matrix $\mat{X}$ are denoted by $\mat{X}_{i*}$, $\mat{X}_{*j}$, and $\mat{X}_{ij}$ respectively. The transpose of matrix $\mat{X}$ is $\mat{X}'$, and its trace is $tr(\mat{X})$ if it is a square matrix. The $\ell_{2}$-norm of the vector $\mat{x}\in\mathbb{R}^{d}$ is denoted by $\|\mat{x}\|_{2}=\sqrt{\sum_{i=1}^{d}x_{i}^{2}}$. The Frobenius norm of matrix $\mat{X}\in\mathbb{R}^{d\times k}$ is $\|\mat{X}\|_{F}=\sqrt{\sum_{i=1}^{d}\sum_{j=1}^{k}\mat{X}_{ij}^{2}}$. The $r$-th power of matrix $\mat{X}$ is denoted as $\mat{X}^{r}$. The main symbols used throughout this paper are summarized in Table~\ref{table:symbols}.

\begin{table}[!t]
\centering
\begin{tabular}{|c|c|} \hline
Notations& Definitions or Descriptions \\ \hline \hline
$G=(\mathcal{V},\mathcal{E})$ & the given network \\ \hline
$n$ & number of nodes in $G$ \\ \hline
$m$ & number of edges in $G$ \\ \hline
$d$ & dimension of the final embedding representation\\ \hline
$d_{s}$ & dimension of each single level embedding  \\ \hline
$k$ & number of levels of the final multi-level embedding \\ \hline
$v_{i}$ & the $i$-th node in $G$\\ \hline
$\mat{A}\in\mathbb{R}_{+}^{n\times n}$ & adjacency matrix of $G$ \\ \hline
$\mat{D}\in\mathbb{R}_{+}^{n\times n}$ & diagonal node degree matrix of $G$ \\ \hline
$\mat{S}=\mat{D}^{-1}\mat{A}$ & transition probability matrix of $G$ \\ \hline
$\mat{U}_{j}\in\mathbb{R}_{+}^{n\times d_{s}}$ & embedding representation of the $j$-th level \\ \hline
$\mat{U}\in\mathbb{R}_{+}^{n\times kd_{s}}$ & the final embedding representation\\ \hline
$\mat{d}$ & degree vector of nodes \\ \hline
$\mat{S}^{k}$ & $k$-step transition probability matrix \\ \hline
$\text{vol}(G)$ & volume of $G$ ($\text{vol}(G)=\sum_{i}\sum_{j}\mat{A}_{ij}$) \\ \hline
\end{tabular}
\caption{Symbols.}
%\vspace{-0.2in}
\label{table:symbols}
\end{table}

Let $G=(\mathcal{V},\mathcal{E})$ be the given network, where $\mathcal{V}=\{v_{1},...,v_{n}\}$ is the node set and $\mathcal{E}=\{e_{1},...,e_{m}\}$ is the edge set. We use the matrix $\mat{A}\in\mathbb{R}^{n\times n}$ to denote the adjacency matrix of the network, where $\mat{A}_{ij}\geq 0$ is a real number denoting the edge weight between node $v_{i}$ and node $v_{j}$. If there is no edge between $v_{i}$ and $v_{j}$, then $\mat{A}_{ij}=0$. In this work, we focus on undirected network such that $\mat{A}_{ij}=\mat{A}_{ji}$ though our problem can be easily extended to directed networks. Now, we formally define the studied problem of multi-level network embedding as follows.
\begin{problem}
(\textbf{Multi-Level Network Embedding}): Given a network $G=(\mathcal{V},\mathcal{E})$ with $n$ nodes, a predefined embedding dimensionality $d$ and the number of levels $k$, the problem of multi-level network embedding is to learn a series of $k$ embeddings $\mat{U}_{i}\in\mathbb{R}^{n\times d_{s}}$ ($i=1,...,k$) for each node in the network ($d=kd_{s} \ll n$). The target is to ensure that the complex node connectivity information in the original network is gradually preserved from coarse to fine in a forward stagewise manner as the levels progress from 1 to $k$.
\end{problem}

\section{The Proposed BoostNE Framework}
\begin{figure*}[!t]
\centering
\includegraphics[width=0.95\textwidth]{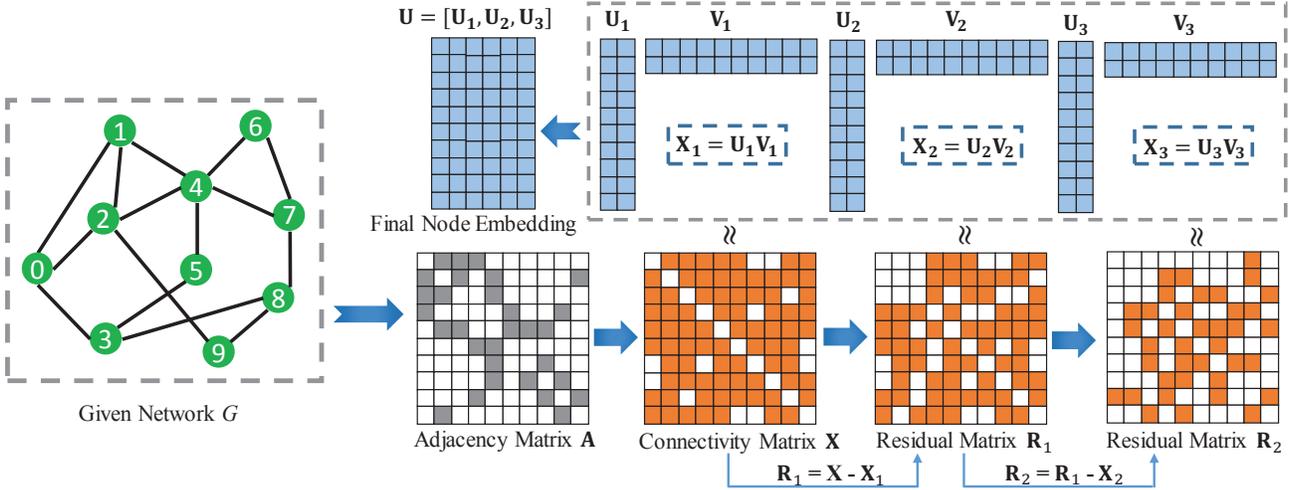}
\caption{An illustration of the proposed framework - BoostNE. BoostNE first constructs the complex node connectivity matrix $\mat{X}$ from the adjacency matrix $\mat{A}$. Motivated by the principle of gradient boosting, the proposed BoostNE successively factorizes the residual of the connectivity matrix from the previous stage and obtains embedding representations from coarse (e.g., $\mat{U}_{1}$) to fine (e.g., $\mat{U}_{3}$). The final embedding representation $\mat{U}$ is obtained by ensembling all previous weak embedding representations.}
\label{fig:framework}
%\vspace{-0.2in}
\end{figure*}

In this section, we present the proposed multi-level network embedding framework with boosted low-rank matrix approximation - BoostNE in details. The key idea is to first formulate the network embedding problem within the framework of matrix factorization by constructing a closed-form node connectivity matrix, then it successively factorizes the residual (unexplained part) obtained from the previous stage. To this end, it enables us to generate a sequence of embedding representations of different granularity views from coarse to fine. Afterwards, we ensemble all these weak embedding representations as the final embedding for downstream learning tasks. An illustrative example of the workflow of the proposed BoostNE framework is shown in Fig.~\ref{fig:framework}.

\subsection{Network Embedding as Matrix Factorization}
As mentioned previously, a large number of existing network embedding methods are fundamentally based on the technique of matrix factorization such as eigendecomposition, singular value decomposition (SVD) and nonnegative matrix factorization (NMF). Concretely, they target at building a deterministic connectivity matrix of various forms to capture different types of connections among nodes in a network, including the first-order proximity, the higher-order proximity, the community structure, and etc. The other category of methods are mostly inspired by the language model word2vec~\cite{mikolov2013distributed}. They either exploit random walks on the network to capture the node structural relations or directly perform edge modeling to learn the structure preserving node representations. Even though their optimization target is distinct from matrix factorization based methods, recent studies~\cite{qiu2018network,yang2015network} examined and found that most of these algorithms can also be reduced to the matrix factorization framework, and the desired embedding representations can be derived by performing principled matrix factorization methods. Next, we briefly show the connections between these sophisticated network embedding methods and matrix factorization.

\noindent{\textbf{Spectral Clustering}~\cite{belkin2002laplacian,tang2011leveraging}:} Given the adjacency matrix $\mat{A}$ and the diagonal degree matrix $\mat{D}$, the node embedding representation $\mat{U}$ can be obtained by concatenating the top-$d$ eigenvectors of the normalized adjacency matrix $\mat{D}^{-1/2}\mat{A}\mat{D}^{-1/2}$.

\noindent{\textbf{Graph Factorization}~\cite{ahmed2013distributed}:} The node embedding $\mat{U}$ is obtained by performing symmetric matrix factorization on the adjacency matrix $\mat{A}$ such that $\mat{U}=\argmin_{\mat{U}}\|\mat{A}-\mat{U}\mat{U}'\|_{F}^{2}$.

\noindent{\textbf{Social Dimension}~\cite{tang2009relational}:} It first constructs the modularity matrix as $\mat{A}-\mat{d}\mat{d}'/2m$, where $\mat{d}$ is the node degree vector. Afterwards, the node embedding $\mat{U}$ can be identified by taking the top-$d$ eigenvectors of the modularity matrix.

\noindent{\textbf{GraRep}~\cite{cao2015grarep}:} It concatenates the top left singular vectors of a series of transition matrices $\mat{X}_{p}$ ($p=1,...,K$). In particular, $K$ denotes the number of transition steps, $\mat{X}_{p}=\text{max}(\mat{Y}_{p},0)$ with $(\mat{Y}_{p})_{ij}=\text{log}(\mat{S}^{p}_{ij}/\sum_{t}\mat{S}^{p}_{tj})-\text{log}(b)$, where $b$ denotes the log shift factor.

\noindent{\textbf{Hope}~\cite{ou2016asymmetric}:} It first derives the connectivity matrix as $\mat{M}=\mat{M}_{g}^{-1}\mat{M}_{l}$ where high-order proximity matrices $\mat{M}_{g}$ and $\mat{M}_{l}$ are built from $\mat{A}$. Then it performs conventional low-rank matrix factorization on $\mat{M}$ to get the node embeddings.

\noindent{\textbf{Deepwalk}~\cite{perozzi2014deepwalk}:} Deepwalk makes an analogy between truncated random walks on networks with sentences in a text corpus and learns the node embedding as the skip-gram model. It has been shown~\cite{qiu2018network} that Deepwalk is equivalent to factorize a transformation of network's normalized Laplacian matrix $\text{log}(\text{vol}(G)(\frac{1}{T}\sum_{r=1}'\mat{S}^{r})\mat{D}^{-1})-\text{log}(b)$, where $\text{vol}(G)$ denotes the volume of the network $G$, $T$ denotes the context window size and $b$ is the number of negative samples.

\noindent{\textbf{node2vec}~\cite{grover2016node2vec}:} It actually factorizes a more complex matrix with strong connections to the stationary distribution and transition probability of second-order random walk. It is also shown in~\cite{qiu2018network} that the factorized matrix is in the following format $\text{log}\big(\frac{\frac{1}{2T}\sum_{r=1}'(\sum_{u}\mat{Y}_{wu}\mat{P}_{cwu}^{r}+\sum_{u}\mat{Y}_{cu}\mat{P}_{wcu}^{r})}{(\sum_{u}\mat{Y}_{wu})(\sum_{u}\mat{Y}_{cu})}\big)-\text{log}(b)$, where $\mat{P}_{uvw}$ denotes the random walk probability to node $u$ given the current visited node $v$ and previously visited node $w$. $\mat{Y}$ encodes the second-order random walk stationary distribution.

\noindent{\textbf{LINE}~\cite{tang2015line}:} LINE is a special case of Deepwalk when the size of context is specified as one~\cite{qiu2018network} and it is equivalent to factorize the following matrix $\text{log}(\text{vol}(G)\mat{S}\mat{D}^{-1})-\text{log}(b)$.

In this following context, we focus on the node connectivity matrix of Deepwalk with the following reasons: (1) it is more general than other node connectivity matrices as it captures both local and global interactions among nodes in the network; (2) it often leads better learning performance as shown in~\cite{qiu2018network}.

\subsection{Multi-Level Network Embedding with Boosted Low-Rank Matrix Approximation}
As per the above summarization, after various node connectivity matrices are derived from the original network topology, a vast majority of existing network embedding methods discover the node embeddings with the prevalent assumption that the underlying connectivity matrix is of low-rank. To this end, they target at learning the node embeddings with low-rank matrix approximation methods by conjecturing that the discriminative connectivity information should be well encoded within a single low-dimensional node representation.

As in the case of Deepwalk, the resultant node connectivity matrix is a nonnegative matrix, and to permit meaningful embedding representations, one possible solution is to perform nonnegative matrix factorization (NMF) on the connectivity matrix $\mat{X}\in\mathbb{R}_{+}^{n\times n}$ with a global low-rank assumption:
\begin{equation}
\min_{\mat{U},\mat{V}\geq 0}\|\mat{X}-\mat{U}\mat{V}\|_{F}^{2},
\label{eq:nmf}
\end{equation}
where $d$ denotes the dimension of the embedding representation. The factorized matrices $\mat{U}\in\mathbb{R}_{+}^{n\times d}$ and $\mat{V}\in\mathbb{R}_{+}^{d\times n}$ are both low-rank ($d\ll n$) and nonnegative. In particular, we can interpret $\mat{U}$ as the embedding of nodes that act like a ``center" node while $\mat{V}'$ can be regarded as the embedding of nodes that play the role of ``context" node~\cite{levy2014neural}. The above objective function of NMF can be solved by conventional optimization algorithms such as coordinate descent~\cite{lee2001algorithms} and projected gradient descent~\cite{lin2007projected}.

In the above optimization problem, in order to perform NMF, the factorized matrix $\mat{X}$ has to satisfy the property of global low-rank. However, it is often argued that the low-rank assumption does not necessarily hold in real-world scenarios, especially when complex interactions among nodes are involved~\cite{lee2013local,yao2015colorization,zhao2017collaborative}. In light of this, performing a single NMF on the connectivity matrix $\mat{X}$ may lead to suboptimal results and the obtained ``one-size fits all" embedding representation is insufficient to encode the connectivity patterns among nodes in the original network.

To find a better embedding representation that well approximates the node connectivity matrix $\mat{X}$, we relax the low-rank assumption of a single run of factorization. Instead, we assume that the connectivity matrix can be well approximated by performing multiple levels of NMF, resulting in the following objective function:
\begin{equation}
\min_{\mat{U}_{i},\mat{V}_{i}\geq 0, i=1,...,k}\|\mat{X}-\sum_{i=1}^{k}\mat{U}_{i}\mat{V}_{i}\|_{F}^{2}.
\end{equation}
In the above objective function, $k$ denotes the number of levels, and $\mat{U}_{i}\in\mathbb{R}_{+}^{n\times d_{s}}$ and $\mat{V}_{i}\in\mathbb{R}_{+}^{d_{s}\times n}$ denotes the embedding representation of the center node and the context node in the $i$-th level during the multi-level factorization process.

In this work, inspired by the well-known ensemble learning methods~\cite{dietterich2000ensemble,friedman2001elements,suh2016ensnmf}, we propose to learn multiple levels of node representations by using a forward stagewise strategy. The essential idea is to perform a sequence of NMF operations with each single NMF operation focusing on the residual of the previously not well approximated part. Hence, the initial embedding representations provide a coarse view of the node connectivity patterns while the latter embeddings provide finer-grained embedding representations. In other words, different stages present various views of the embedding of different granularity. More specifically, in the $i$-th level, the not well explained residual matrix is defined as follows:
\begin{align}
\mat{R}_{i}=\left\{
\begin{array}{ll}
\mat{X}& \,\,\, \text{if         }    i=1 \\
\text{max}(\mat{R}_{i-1}-\mat{U}_{i-1}\mat{V}_{i-1},0)& \,\,\, \text{if         }    i\geq2.\\
\end{array}
\right.
\label{eq:residual}
\end{align}
The max operation implies that if there exists any negative elements after the approximation, we convert it to be zero. With the above defined residual matrix, the embedding representation at the $i$-the level is obtained by minimizing the following loss function:
\begin{equation}
\min_{\mat{U}_{i},\mat{V}_{i}\geq 0}  \|\mat{R}_{i}-\mat{U}_{i}\mat{V}_{i}\|_{F}^{2}.
\label{eq:nmfsub}
\end{equation}

Compared with the objective function in Eq.~(\ref{eq:nmf}) which returns the node embedding representation in a single run with NMF, our proposed algorithm returns multiple weak representations from coarse to fine in a greedy fashion. In other words, once the earlier level embeddings $\mat{U}_{i}$ and $\mat{V}_{i}'$ are obtained, they are fixed for the remaining operations. Next, we briefly analyze the time complexity of the proposed BoostNE and NMF with a single run in Eq.~(\ref{eq:nmf}), here we specify $d=d_{s}k$ for a fair comparison. The time complexity of optimizing Eq.~(\ref{eq:nmf}) is related to the number of nonzero elements in $\mat{X}$ and the rank $d$, which is $\#iterations\times O(\text{nnz}(\mat{X})d)$. While for the proposed BoostNE, as can be observed from the illustrative example in Figure~\ref{fig:framework}, the residual matrix (the unexplained part) becomes sparser and sparser as the level goes up, thus the computational cost of the proposed BoostNE is $\#iterations\times O(\text{nnz}(\mat{R}_{1}+,...,+\mat{R}_{k})d_{s})$. As $\text{nnz}(\mat{R}_{1})+,...,+\text{nnz}(\mat{R}_{k})<\text{nnz}(\mat{X})k$, the proposed BoostNE is more efficient than performing a single run NMF in obtaining a ``one-size fits all" embedding representation. The detailed pseudo code of the proposed multi-level network embedding framework BoostNE is shown in Algorithm~\ref{alg:boostne}.

\begin{algorithm}[!t]
\normalsize
\begin{algorithmic}[1]
    \REQUIRE  A given network $G=(\mathcal{V},\mathcal{E})$; number of levels $k$, embedding dimension in each level $d_{s}$.
    \ENSURE The node embedding representation $\mat{U}_{1},...,\mat{U}_{k}$ from coarse to fine and the final embedding representation $\mat{U}$.
    \STATE {Obtain the node connectivity matrix $\mat{X}$ from the adjacency matrix $\mat{A}$ of the network;}
    \FOR {$i = 1$ to $k$}
        \STATE {Compute the residual matrix $\mat{R}_{i}$ with Eq.~(\ref{eq:residual});}
        \STATE {Obtain the center node embedding representation $\mat{U}_{i}$ and the context node embedding representation $\mat{V}_{i}'$ by alternating optimization algorithms;}
    \ENDFOR
    \STATE{Return the final embedding as $\mat{U}=[\mat{U}_{1},...,\mat{U}_{k}]$}
\end{algorithmic}
\caption{The proposed multi-level network embedding framework - BoostNE}
\label{alg:boostne}
\end{algorithm}

\section{Experiments}
In this section, we conduct experiments to assess the effectiveness of the proposed BoostNE on the task of multi-label node classification. Before introducing the detailed experimental results, we first introduce the used datasets, compared baseline methods, and experimental settings. In this section, we also perform further analysis on why the multi-level embedding methods achieve better performance by studying the approximation error of the connectivity matrix. At last, we investigate how the number of levels $k$ affects the final network embedding results.

\subsection{Dataset}
We collect and use four real-world network datasets from different domains for experimental evaluation. All these four datasets are publicly available and have been used in previous research~\cite{perozzi2014deepwalk,yang2015network,grover2016node2vec}. The detailed descriptions of these datasets are as follow:
\begin{itemize}
\item \textbf{Cora}\footnote{https://linqs.soe.ucsc.edu/data}: Cora is a citation network of scientific publications. It consists of 2,708 papers from 7 classes representing different research areas. There is a total number of 5,278 citation links in the dataset.
\item \textbf{Wiki}\footnote{https://github.com/thunlp/OpenNE/tree/master/data/wiki}: Wiki is a collection of wikipedia documents that are inherently connected with each other via hyperlinks. Each document is categorized into a number of predefined classes denoting their topics. In total, we have 2,363 documents, 11,596 hyperlinks and 17 topics.
\item \textbf{PPI}\footnote{https://snap.stanford.edu/node2vec/\#datasets}: It is a subgraph of the protein-protein interaction network of Homo Sapiens. In the dataset, the labels of the protein are obtained from the hallmark gene sets and denote the biological states. We have 3,860 proteins, 37,845 interaction links, and 50 states.
\item \textbf{Blogcatalog}\footnote{http://socialcomputing.asu.edu/datasets/BlogCatalog} is a social blogging website in which users follow each other and post blogs under certain predefined categories. The main categories of blogs by the users are regarded as the class labels of users. In the dataset, we have 10,312 users, 333,983 user relations and 39 predefined categories.
\end{itemize}
For all the above mentioned datasets, we have removed all self-loop edges and have converted bi-directional edges to undirected ones for a fair comparison of various network embedding methods. The detailed statistics of these datasets are listed in Table~\ref{tab:statistics}.

\begin{table}[!t]
\small
\begin{center}
\begin{tabular}{c|c|c|c|c }\hline
\textbf{Dataset} & Cora & Wiki & PPI & Blogcatalog  \\ \hline \hline
$|\mathcal{V}|$ & 2,708 & 2,363 & 3,860 & 10,312 \\ \hline
$|\mathcal{E}|$ & 5,278 & 11,596 & 37,845 & 333,983 \\ \hline
$\#$ of labels & 7 & 17 & 50 & 39 \\ \hline
\end{tabular}
\end{center}
\caption{Statistics of the used datasets.}
\vspace{-0.1in}
\label{tab:statistics}
\end{table}

\subsection{Compared Baseline Methods}
In this subsection, we compare our proposed multi-level network embedding framework BoostNE with existing efforts from two main categories: matrix factorization based network embedding methods and skip-gram inspired network embedding methods. Among them, Spectral Clustering, Social Dimension, GraRep belong to the former category while Deepwalk and LINE fall into the latter one. In addition, we also compare with a recently proposed framework NetMF which provides a general framework to factorize a closed-form node connectivity matrix for embedding learning.
\begin{itemize}
\item \textbf{Spectral Clustering}~\cite{tang2011leveraging}: It is a typical matrix factorization based approach which takes the top-$d$ eigenvectors the normalized Laplacian matrix of network $G$ as the node embedding representation.
\item \textbf{Modularity}~\cite{tang2009relational}: It is a kind of matrix factorization based method by taking the top-$d$ eigenvectors of the modularity matrix from the network $G$ why assuming that good embedding assignment can maximize the modularity of the node partition.
\item \textbf{GraRep}~\cite{cao2015grarep}: This method learns node embedding representation by capturing different $k$-step relational information among nodes. This method also generates multiple node embedding representations. However, it is different from our method as they operate on multiple transitional matrices while our method focuses on a single node connectivity matrix.
\item \textbf{Deepwalk}~\cite{perozzi2014deepwalk}: Deepwalk is the first network embedding method which borrows the idea of word2vec in the NLP community. Specifically, it performs truncated random walks on the network and the node embeddings are learned by capturing the node proximity information encoded in these short random walks.
\item \textbf{LINE}~\cite{tang2015line}: LINE carefully designs the objective function in preserving the first-order and the second-order node proximity for node embedding representation learning. We concatenate the embedding representations from these two objective functions together.
\item \textbf{NetMF}~\cite{qiu2018network}: It is a recently proposed network embedding method which bridges the gap between matrix factorization based approaches and skip-gram inspired methods. In particular, it attempts to approximate the closed-form of the Deepwalk's implicit matrix. In this work, to have a fair comparison with our method, we adapt it by performing NMF instead of SVD after the closed-form connectivity matrix is derived.
\end{itemize}

\begin{table*}[!t]
\centering
\begin{tabular}{|c|c||c|c|c|c|c|c|c|c|c|}
\hline
\multicolumn{2}{|c||}{Training Ratio} & 10\% & 20\% & 30\% & 40\% & 50\% & 60\% & 70\% & 80\% & 90\%  \\ \hline \hline
\multirow{6}{*}{Micro-F1}
			& Spectral Clustering  & 0.6814 & 0.7597 & 0.7777 & 0.7775 & 0.8091 & 0.7986 & 0.8058 & 0.8157 & 0.8170  \\ \cline{2-11}
			& Social Dimension     & 0.5459	& 0.6604 & 0.7004 & 0.7252 & 0.7423	& 0.7605 & 0.7635 & 0.7832 & 0.7789  \\ \cline{2-11}
			& GraRep               & 0.7636	& 0.7774 & 0.7818 & 0.7818 & 0.7887	& 0.7864 & 0.7861 & 0.7954 & 0.7815  \\ \cline{2-11}
			& Deepwalk             & 0.7589	& 0.7874 & 0.7989 & 0.8039 & 0.8151 & 0.8181 & 0.8118 & 0.8175 & 0.8162  \\ \cline{2-11}
			& LINE                 & 0.6588	& 0.7126 & 0.7271 & 0.7439 & 0.7514	& 0.7598 & 0.7700 & 0.7655 & 0.7609 \\ \cline{2-11}
			& NetMF                & 0.7379	& 0.7829 & 0.8002 & 0.8072 & 0.8120	& 0.8152 & 0.8180 & 0.8275 & 0.8306  \\ \cline{2-11}
			& BoostNE              & \cellcolor{blue!25}0.7824	& \cellcolor{blue!25}0.8047 & \cellcolor{blue!25}0.8178 & \cellcolor{blue!25}0.8250 & \cellcolor{blue!25}0.8257	& \cellcolor{blue!25}0.8266 & \cellcolor{blue!25}0.8314 & \cellcolor{blue!25}0.8367 & \cellcolor{blue!25}0.8373  \\ \hline\hline
			\multirow{6}{*}{Macro-F1}
			& Spectral Clustering  & 0.6746	& 0.7552 & 0.7699 & 0.7764 & 0.8053	& 0.7928 & 0.8004 & 0.8104 & 0.8071  \\ \cline{2-11}
			& Social Dimension     & 0.5032	& 0.6331 & 0.6860 & 0.7130 & 0.7247	& 0.7532 & 0.7540 & 0.7742 & 0.7705  \\ \cline{2-11}
			& GraRep               & 0.7543	& 0.7656 & 0.7710 & 0.7721 & 0.7797	& 0.7778 & 0.7768 & 0.7833 & 0.7720  \\ \cline{2-11}
			& Deepwalk             & 0.7457	& 0.7776 & 0.7886 & 0.7973 & 0.8074	& 0.8094 & 0.8047 & 0.8133 & 0.8049  \\ \cline{2-11}
			& LINE                 & 0.6409	& 0.7007 & 0.7158 & 0.7319 & 0.7423	& 0.7513 & 0.7599 & 0.7564 & 0.7545 \\ \cline{2-11}
			& NetMF                & 0.7220	& 0.7706 & 0.7902 & 0.7977 & 0.8025	& 0.8063 & 0.8109 & 0.8229 & 0.8264  \\ \cline{2-11}
			& BoostNE              & \cellcolor{blue!25}0.7638	& \cellcolor{blue!25}0.7906 & \cellcolor{blue!25}0.8062 & \cellcolor{blue!25}0.8142 & \cellcolor{blue!25}0.8143	& \cellcolor{blue!25}0.8171 & \cellcolor{blue!25}0.8232 & \cellcolor{blue!25}0.8313 & \cellcolor{blue!25}0.8309  \\ \hline
\end{tabular}
\caption{Node classification performance comparison on the Cora dataset.}
\vspace{-0.05in}
\label{table:cora}
\end{table*}

\begin{table*}[!t]
\centering
\begin{tabular}{|c|c||c|c|c|c|c|c|c|c|c|}
\hline
\multicolumn{2}{|c||}{Training Ratio} & 10\% & 20\% & 30\% & 40\% & 50\% & 60\% & 70\% & 80\% & 90\%  \\ \hline \hline
\multirow{6}{*}{Micro-F1}
			& Spectral Clustering  & 0.5333	& 0.5761 & 0.6236 & 0.6353 & 0.6460	& 0.6432 & 0.6597 & 0.6432 & 0.6419  \\ \cline{2-11}
			& Social Dimension     & 0.3920	& 0.5054 & 0.5727 & 0.5991 & 0.6182	& 0.6289 & 0.6386 & 0.6282 & 0.6496  \\ \cline{2-11}
			& GraRep               & 0.5948	& 0.6078 & 0.6127 & 0.6242 & 0.6338	& 0.6354 & 0.6329 & 0.6404 & 0.6346  \\ \cline{2-11}
			& Deepwalk             & 0.5935	& 0.6324 & 0.6485 & 0.6606 & 0.6654	& \cellcolor{blue!25}0.6865 & 0.6732 & 0.6907 & 0.6713  \\ \cline{2-11}
			& LINE                 & 0.5609	& 0.6032 & 0.6283 & 0.6356 & 0.6552	& 0.6659 & 0.6729 & 0.6732 & 0.6768 \\ \cline{2-11}
			& NetMF                & 0.5849	& 0.6306 & 0.6508 & 0.6642 & 0.6709	& 0.6786 & 0.6817 & 0.6848 & 0.6916  \\ \cline{2-11}
			& BoostNE              & \cellcolor{blue!25}0.6113	& \cellcolor{blue!25}0.6442 & \cellcolor{blue!25}0.6625 & \cellcolor{blue!25}0.6709 & \cellcolor{blue!25}0.6749	& 0.6846 & \cellcolor{blue!25}0.6829 & \cellcolor{blue!25}0.6915 & \cellcolor{blue!25}0.7013  \\ \hline\hline
			\multirow{6}{*}{Macro-F1}
			& Spectral Clustering  & 0.4012	& 0.4459 & 0.4774 & 0.4959 & 0.5107	& 0.5111 & 0.5364 & 0.5119 & 0.4952  \\ \cline{2-11}
			& Social Dimension     & 0.3154	& 0.4082 & 0.4584 & 0.4802 & 0.4946	& 0.5013 & 0.5036 & 0.5021 & 0.5028  \\ \cline{2-11}
			& GraRep               & 0.4092	& 0.4244 & 0.4220 & 0.4399 & 0.4557 & 0.4590 & 0.4528 & 0.4517 & 0.4619  \\ \cline{2-11}
			& Deepwalk             & 0.4368	& \cellcolor{blue!25}0.4907 & 0.5051 & \cellcolor{blue!25}0.5347 & 0.5353	& 0.5475 & 0.5531 & \cellcolor{blue!25}0.5695 & 0.5492  \\ \cline{2-11}
			& LINE                 & 0.4314	& 0.4655 & 0.4917 & 0.5088 & 0.5448	& 0.5359 & 0.5409 & 0.5560 & 0.5334 \\ \cline{2-11}
			& NetMF                & 0.4188	& 0.4595 & 0.4797 & 0.4998 & 0.5192	& 0.5304 & 0.5364 & 0.5391 & 0.5455  \\ \cline{2-11}
			& BoostNE              & \cellcolor{blue!25}0.4421	& 0.4742 & \cellcolor{blue!25}0.5055 & 0.5248 & \cellcolor{blue!25}0.5404	& \cellcolor{blue!25}0.5496 & \cellcolor{blue!25}0.5562 & 0.5645 & \cellcolor{blue!25}0.5692  \\ \hline
\end{tabular}
\caption{Node classification performance comparison on the Wiki dataset.}
\vspace{-0.05in}
\label{table:wiki}
\end{table*}

\begin{table*}[!t]
\centering
\begin{tabular}{|c|c||c|c|c|c|c|c|c|c|c|}
\hline
\multicolumn{2}{|c||}{Training Ratio} & 10\% & 20\% & 30\% & 40\% & 50\% & 60\% & 70\% & 80\% & 90\%  \\ \hline \hline
\multirow{6}{*}{Micro-F1}
			& Spectral Clustering  & 0.1629 & 0.1881 & 0.1942 & 0.2027 & 0.2085 & 0.2108 & 0.2131 & 0.2177 & 0.2086  \\ \cline{2-11}
			& Social Dimension     & 0.1250	& 0.1399 & 0.1622 & 0.1798 & 0.1795	& 0.1880 & 0.1945 & 0.1985 & 0.1966  \\ \cline{2-11}
			& GraRep               & 0.1828	& 0.2034 & 0.2121 & 0.2183 & 0.2239 & 0.2278 & 0.2280 & 0.2313 & 0.2292  \\ \cline{2-11}
			& Deepwalk             & 0.1588	& 0.1754 & 0.1889 & 0.1987 & 0.2075	& 0.2075 & 0.2155 & 0.2234 & 0.2238  \\ \cline{2-11}
			& LINE                 & 0.1342	& 0.1548 & 0.1706 & 0.1811 & 0.1923	& 0.2014 & 0.2063 & 0.2096 & 0.2180 \\ \cline{2-11}
			& NetMF                & 0.1771	& 0.2033 & 0.2171 & 0.2264 & 0.2329 & 0.2401 & 0.2466 & 0.2531 & 0.2565  \\ \cline{2-11}
			& BoostNE              & \cellcolor{blue!25}0.1883	& \cellcolor{blue!25}0.2137 & \cellcolor{blue!25}0.2271 & \cellcolor{blue!25}0.2367 & \cellcolor{blue!25}0.2430 & \cellcolor{blue!25}0.2487 & \cellcolor{blue!25}0.2534 & \cellcolor{blue!25}0.2580 & \cellcolor{blue!25}0.2620  \\ \hline\hline
			\multirow{6}{*}{Macro-F1}
			& Spectral Clustering  & 0.1123	& 0.1320 & 0.1412 & 0.1439 & 0.1526 & 0.1582 & 0.1571 & 0.1588 & 0.1497  \\ \cline{2-11}
			& Social Dimension     & 0.1001	& 0.1170 & 0.1326 & 0.1508 & 0.1520	& 0.1576 & 0.1639 & 0.1588 & 0.1626 \\ \cline{2-11}
			& GraRep               & 0.1432	& 0.1613 & 0.1697 & 0.1736 & 0.1809	& 0.1842 & 0.1854 & 0.1847 & 0.1856  \\ \cline{2-11}
			& Deepwalk             & 0.1282	& 0.1483 & 0.1622 & 0.1729 & 0.1781	& 0.1772 & 0.1863 & 0.1874 & 0.1843  \\ \cline{2-11}
			& LINE                 & 0.1107	& 0.1331 & 0.1467 & 0.1559 & 0.1672 & 0.1744 & 0.1767 & 0.1771 & 0.1828 \\ \cline{2-11}
			& NetMF                & 0.1384	& 0.1662 & 0.1800 & 0.1892 & 0.1950	& 0.2005 & 0.2062 & \cellcolor{blue!25}0.2101 & \cellcolor{blue!25}0.2109  \\ \cline{2-11}
			& BoostNE              & \cellcolor{blue!25}0.1469	& \cellcolor{blue!25}0.1726 & \cellcolor{blue!25}0.1867 & \cellcolor{blue!25}0.1971 & \cellcolor{blue!25}0.2011 & \cellcolor{blue!25}0.2035 & \cellcolor{blue!25}0.2082 & 0.2098 & 0.2098  \\ \hline
\end{tabular}
\caption{Node classification performance comparison on the PPI dataset.}
\vspace{-0.05in}
\label{table:ppi}
\end{table*}

\begin{table*}[!t]
\centering
\begin{tabular}{|c|c||c|c|c|c|c|c|c|c|c|}
\hline
\multicolumn{2}{|c||}{Training Ratio} & 10\% & 20\% & 30\% & 40\% & 50\% & 60\% & 70\% & 80\% & 90\%  \\ \hline \hline
\multirow{6}{*}{Micro-F1}
			& Spectral Clustering  & 0.3663	& 0.3879 & 0.3977 & 0.4044 & 0.4119 & 0.4109 & 0.4140 & 0.4180 & 0.4191  \\ \cline{2-11}
			& Social Dimension     & 0.2721	& 0.3094 & 0.3215 & 0.3324 & 0.3326	& 0.3381 & 0.3411 & 0.3377 & 0.3360  \\ \cline{2-11}
			& GraRep               & 0.3325	& 0.3473 & 0.3553 & 0.3608 & 0.3640	& 0.3701 & 0.3715 & 0.3718 & 0.3712  \\ \cline{2-11}
			& Deepwalk             & 0.3358	& 0.3621 & 0.3777 & 0.3865 & 0.3931	& 0.3991 & 0.4042 & 0.4051 & 0.4171  \\ \cline{2-11}
			& LINE                 & 0.2960	& 0.3336 & 0.3557 & 0.3665 & 0.3755	& 0.3828 & 0.3829 & 0.3898 & 0.3938 \\ \cline{2-11}
			& NetMF                & 0.3752	& 0.3945 & 0.4051 & 0.4133 & 0.4178	& 0.4202 & 0.4235 & 0.4267 & 0.4305  \\ \cline{2-11}
			& BoostNE              & \cellcolor{blue!25}0.3908	& \cellcolor{blue!25}0.4088 & \cellcolor{blue!25}0.4173 & \cellcolor{blue!25}0.4242 & \cellcolor{blue!25}0.4291	& \cellcolor{blue!25}0.4304 & \cellcolor{blue!25}0.4322 & \cellcolor{blue!25}0.4364 & \cellcolor{blue!25}0.4410  \\ \hline\hline
			\multirow{6}{*}{Macro-F1}
			& Spectral Clustering  & 0.2142	& 0.2344 & 0.2450 & 0.2519 & 0.2587 & 0.2573 & 0.2612 & 0.2608 & 0.2610  \\ \cline{2-11}
			& Social Dimension     & 0.1503	& 0.1650 & 0.1757 & 0.1805 & 0.1808	& 0.1818 & 0.1841 & 0.1766 & 0.1768  \\ \cline{2-11}
			& GraRep               & 0.1625	& 0.1787 & 0.1851 & 0.1899 & 0.1943 & 0.1976 & 0.1977 & 0.1998 & 0.2003  \\ \cline{2-11}
			& Deepwalk             & 0.1904	& 0.2186 & 0.2347 & 0.2440 & 0.2520	& 0.2581 & 0.2677 & 0.2701 & 0.2684  \\ \cline{2-11}
			& LINE                 & 0.1738	& 0.2003 & 0.2176 & 0.2264 & 0.2363 & 0.2433 & 0.2391 & 0.2508 & 0.2532 \\ \cline{2-11}
			& NetMF                & 0.2165	& 0.2358 & 0.2497 & 0.2572 & 0.2615 & 0.2644 & 0.2675 & 0.2686 & 0.2692  \\ \cline{2-11}
			& BoostNE              & \cellcolor{blue!25}0.2325	& \cellcolor{blue!25}0.2544 & \cellcolor{blue!25}0.2674 & \cellcolor{blue!25}0.2759 & \cellcolor{blue!25}0.2807 & \cellcolor{blue!25}0.2844 & \cellcolor{blue!25}0.2851 &\cellcolor{blue!25} 0.2860 &\cellcolor{blue!25} 0.2852  \\ \hline
\end{tabular}
\caption{Node classification performance comparison on the Blogcatalog dataset.}
\vspace{-0.1in}
\label{table:blogcatalog}
\end{table*}

The parameter settings of different embedding algorithms are as follows. For all the compared network embedding methods, we follow~\cite{perozzi2014deepwalk,grover2016node2vec,qiu2018network} to specify the final embedding dimension as 128. For GraRep, the log shit factor $b$ is set as $1/n$, the transition step is specified as 8. For Deepwalk, we set the number of walks as 10, the walk length as 80, and the window size as 10. In terms of LINE, we concatenate both the first-order and the second-order embedding representations, the negative sample size is 5 and the number of training epoches is 100. Lastly, for NetMF, we use the Deepwalk's implicit matrix by following the same setting as Deepwalk, and the parameter $h$ in NetMF is set as 256. For the proposed BoostNE, the only additional parameter we need to specify is the number of levels $k$, we set it as 8. More discussions on the impact of $k$ will be presented later.

\subsection{Experimental Settings}
To assess the effectiveness of the proposed multi-level network embedding framework BoostNE, we follow the commonly adopted setting to compare different embedding algorithms on the task of multi-label node classification. In the task of multi-label node classification, each node is associated with multiple class labels, our target is to build a predictive learning model to predict the correct labels of nodes. In particular, after the embedding representations of all $n$ nodes are obtained, we randomly sample a portion of nodes as the training data, use their embeddings and their class labels to build a predictive classification model, and then make the predictions with the embeddings of the remaining nodes. In the experiments, we repeat the process 10 times and report the average classification performance in terms of Micro-F1 and Macro-F1, which are widely used multi-label classification evaluation metrics. Specifically, Micro-F1 is a weighted average of F1-scores over different classes while Macro-F1 is an arithmetic average of F1-scores from different labels:
\begin{equation}
\begin{split}
\text{Micro-F1}&=\frac{\sum_{i=1}^{c}\text{2TP}^{i}}{\sum_{i=1}^{c}(\text{2TP}^{i}+\text{FP}^{i}+\text{FN}^{i})}\\
\text{Macro-F1}&=\frac{1}{c}\sum_{i=1}^{c}\frac{\text{2TP}^{i}}{(2\text{TP}^{i}+\text{FP}^{i}+\text{FN}^{i})},
\end{split}
\end{equation}
where $\text{TP}^{i}$, $\text{FP}^{i}$, and $\text{FN}^{i}$ denote the number of positives, false positives, and false negatives in the $i$-th class, respectively. Normally, higher values imply better classification performance, which further indicate better node embedding representations. In the experiments, we vary the percentage of training data from 10\% to 90\%, and the logistic regression in Liblinear~\cite{fan2008liblinear} is used as the discriminative classifier.

\begin{figure}[!t]
\centering
\includegraphics[width=0.48\textwidth]{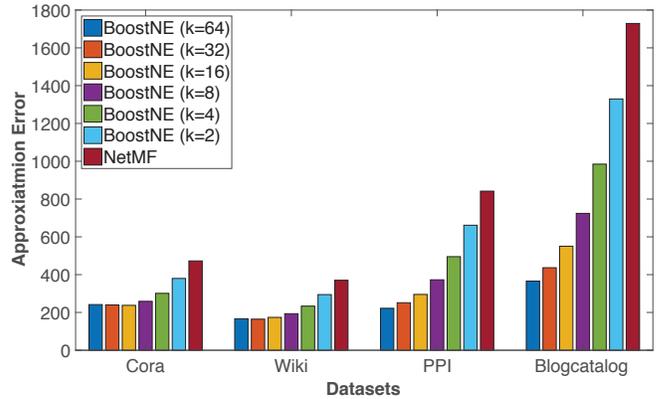}
\caption{Approximation error of the node connectivity of NetMF and BoostNE with different number of levels $k$.}
\label{fig:error}
\vspace{-0.1in}
\end{figure}
\subsection{Embedding Results Comparison}
We first compare the quality of embeddings from different methods in the task of multi-label node classification. The comparison results on the Cora, Wiki, PPI and Blogcatalog datasets are shown in Table~\ref{table:cora}, Table~\ref{table:wiki}, Table~\ref{table:ppi} and Table~\ref{table:blogcatalog}, respectively. We make the following observations from these tables:
\begin{itemize}
\item For all methods, the multi-label node classification performance w.r.t. Micro-F1 and Macro-F1 gradually increases when the portion of training data is varied from 10\% to 90\%. It implies that more training data can help us learn more effective embedding representations.
\item The proposed multi-level network embedding method BoostNE achieves the best classification performance in most of the cases. To further validate the conclusion, we perform a pairwise Wilcoxon signed rank test between BoostNE and other embedding methods and the results indicate that BoostNE is significantly better when the $p$-value threshold is set as 0.05.
\item Compared with NetMF which learns node embedding within a single run of low-rank matrix approximation, the proposed BoostNE learns multiple embedding representations from coarse to fine by gradually factorizing the residual matrix from previous stage. The improvement of BoostNE over NetMF shows the effectiveness of the ensemble methods as multiple weak embedding representations lead to a more effective embedding.
\item Most of the time, the skip-gram inspired methods such as Deepwalk and LINE are better than conventional matrix factorization based methods. Even though both NetMF and BoostNE are kind of matrix factorization based methods, their performance is much better than conventional matrix factorization methods as the connectivity matrix they factorize encode complex node interactions.
\end{itemize}
\begin{figure*}[!ht]
\centering
\begin{minipage}{0.36\textwidth}
\subfigure[Cora]
{\includegraphics[width=\textwidth]{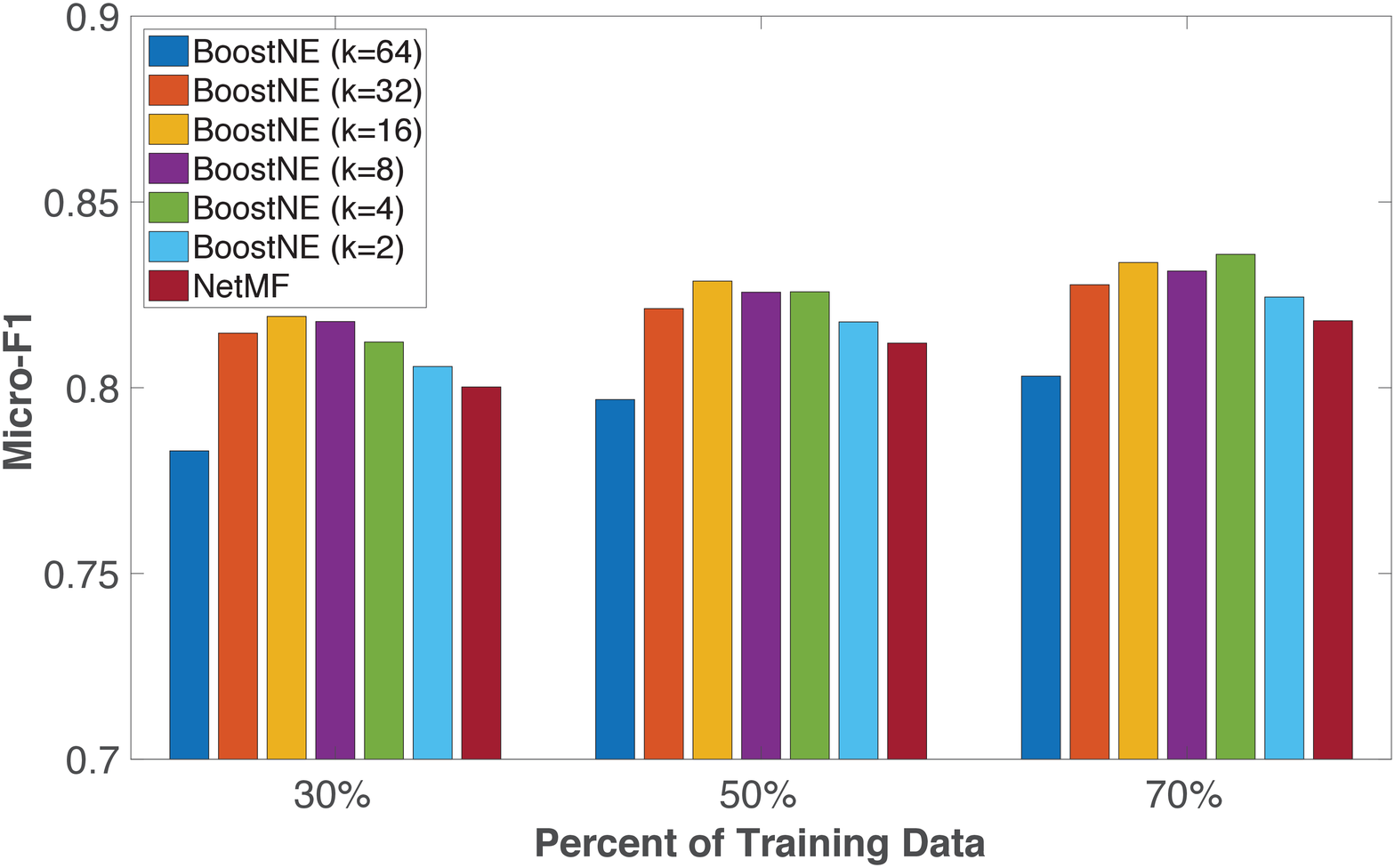}}
\end{minipage}
\hspace{0.02\textwidth}
\begin{minipage}{0.36\textwidth}
\subfigure[Wiki]
{\includegraphics[width=\textwidth]{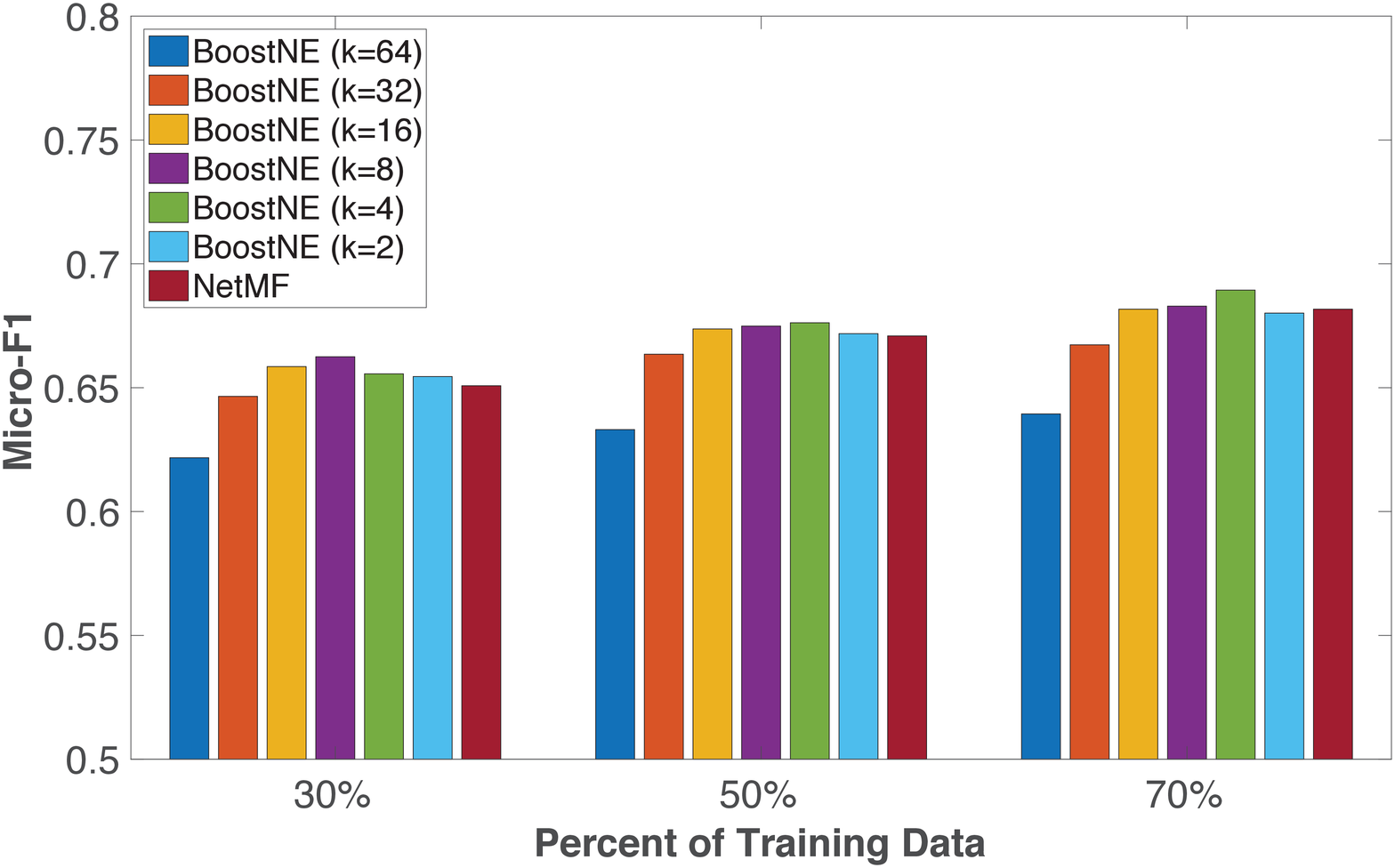}}
\end{minipage}
\begin{minipage}{0.36\textwidth}
\subfigure[PPI]
{\includegraphics[width=\textwidth]{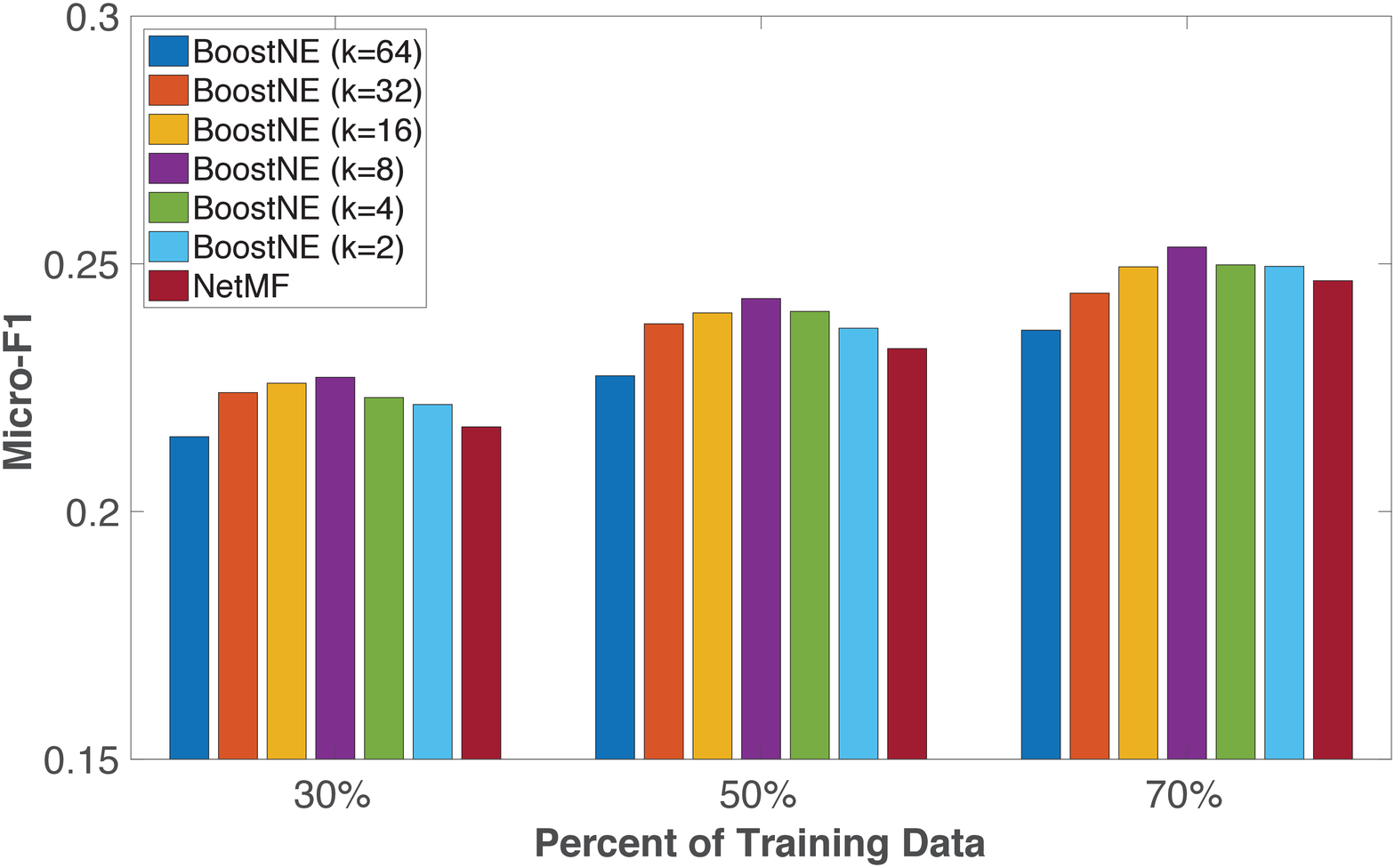}}
\end{minipage}
\hspace{0.02\textwidth}
\begin{minipage}{0.36\textwidth}
\subfigure[Blogcatalog]
{\includegraphics[width=\textwidth]{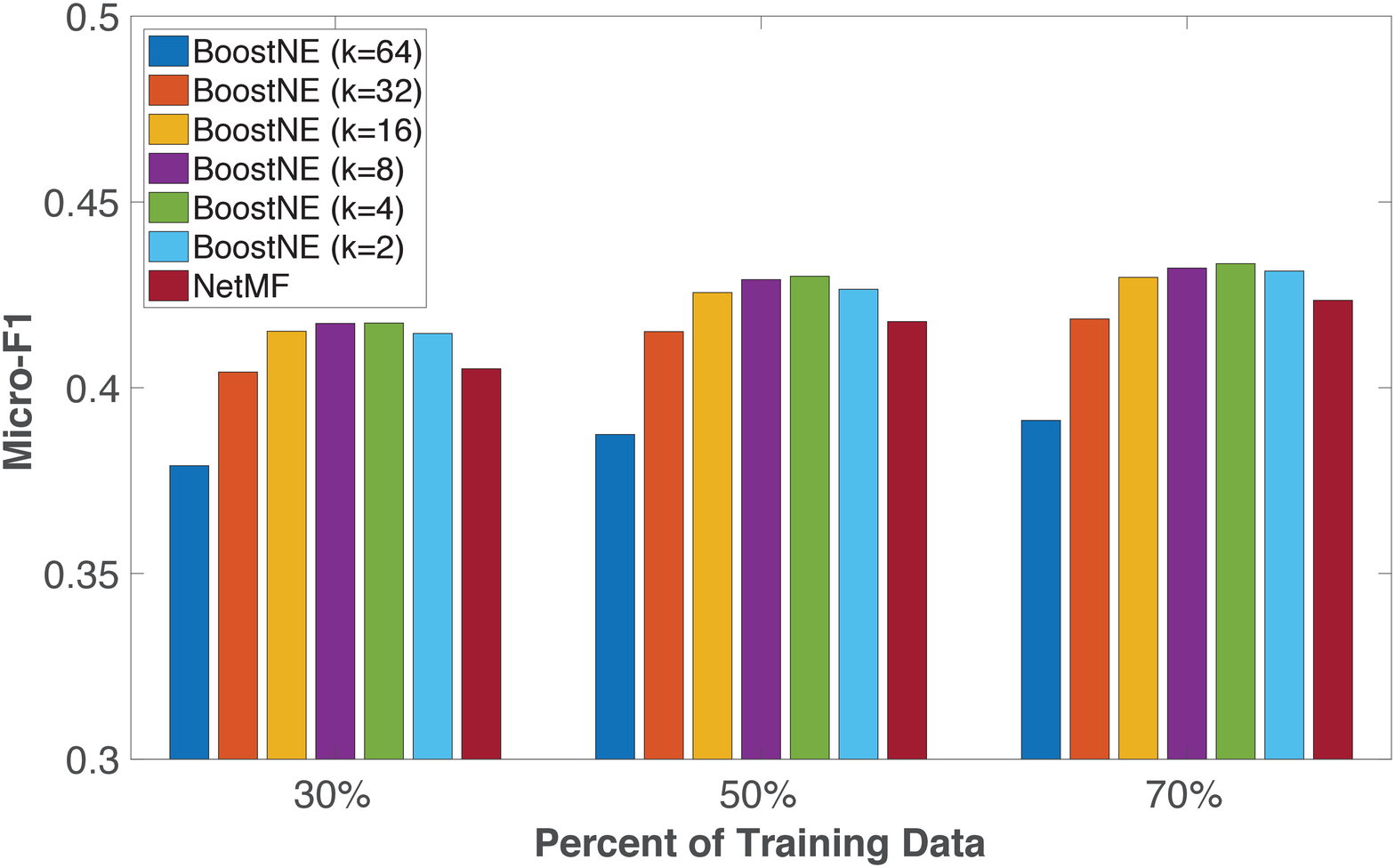}}
\end{minipage}
\centering
\vspace{-0.1in}
\caption{Impact of the number of levels $k$ on the learned node embedding w.r.t. Micro-F1.}
\label{fig:kmicro}
\end{figure*}
\vspace{-0.1in}
\begin{figure*}[!ht]
\centering
\begin{minipage}{0.36\textwidth}
\subfigure[Cora]
{\includegraphics[width=\textwidth]{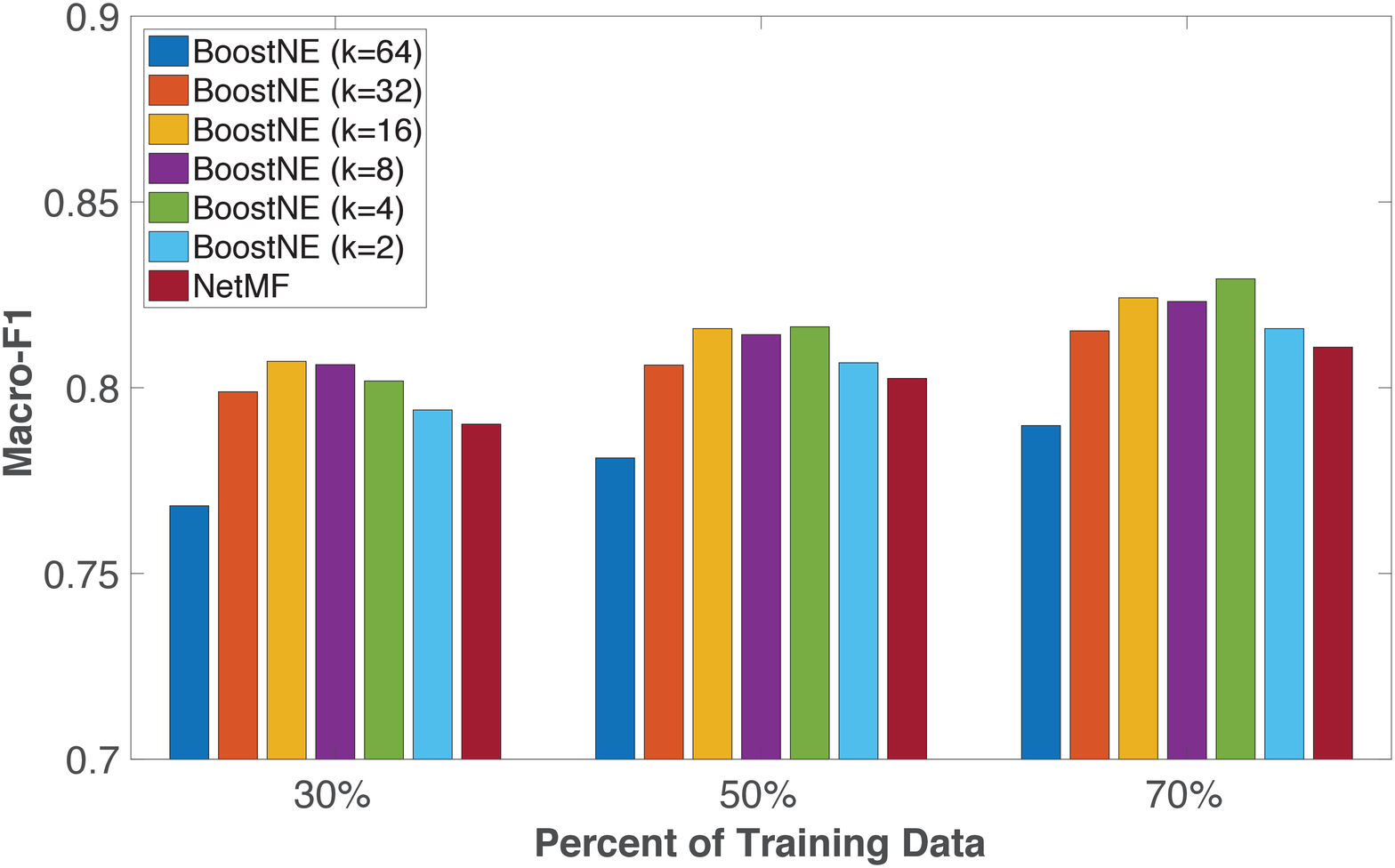}}
\end{minipage}
\hspace{0.02\textwidth}
\begin{minipage}{0.36\textwidth}
\subfigure[Wiki]
{\includegraphics[width=\textwidth]{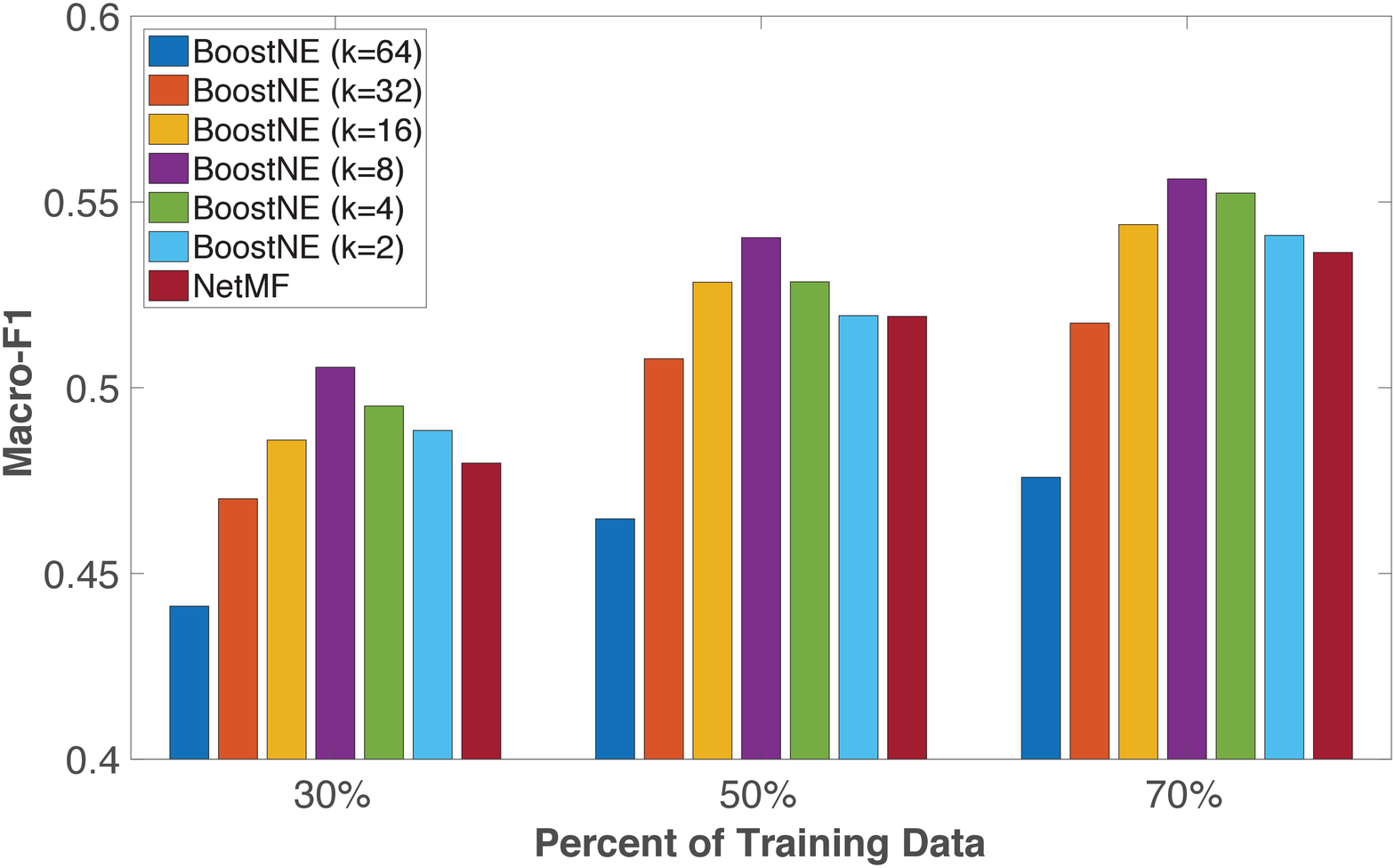}}
\end{minipage}
\begin{minipage}{0.36\textwidth}
\subfigure[PPI]
{\includegraphics[width=\textwidth]{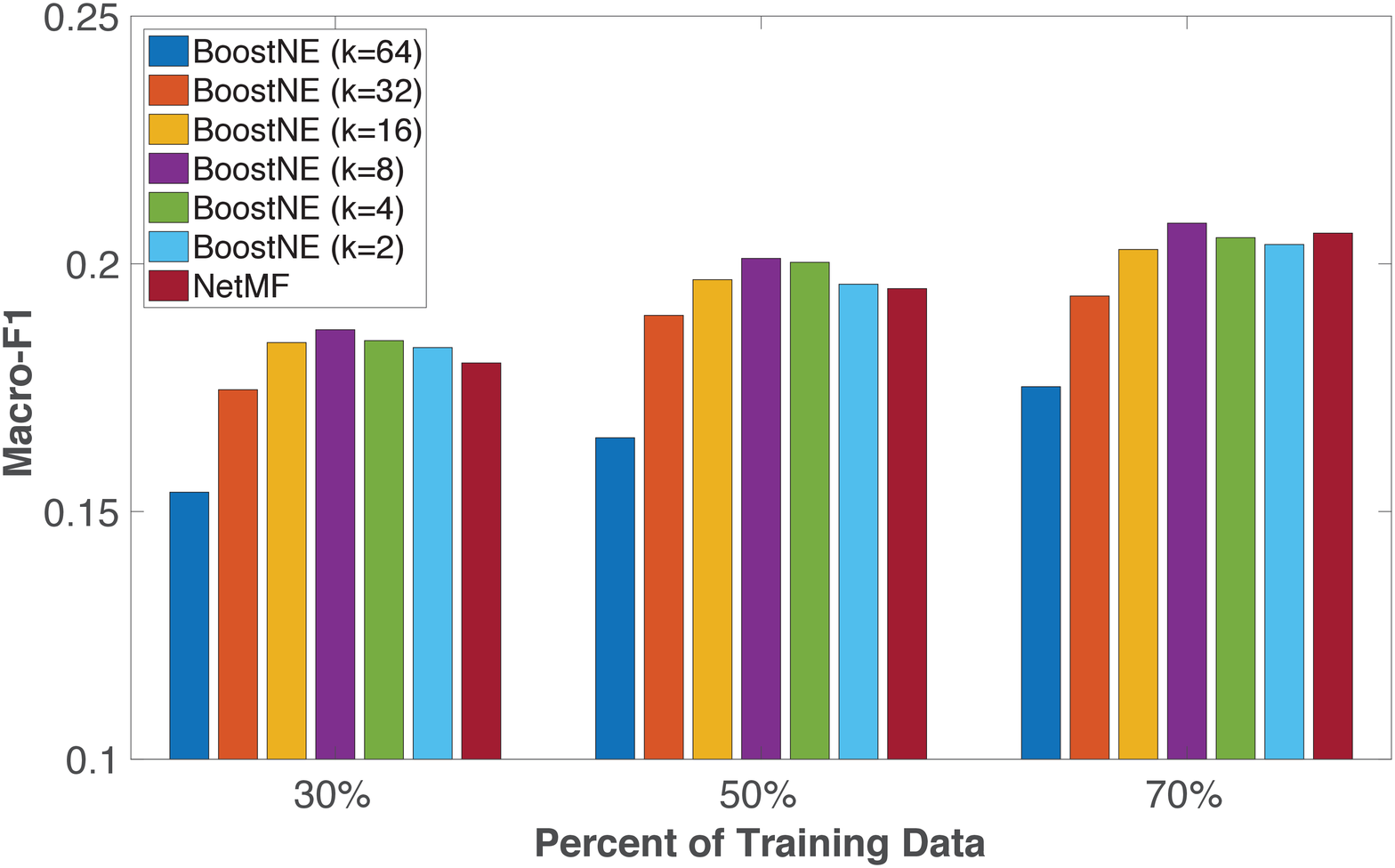}}
\end{minipage}
\hspace{0.02\textwidth}
\begin{minipage}{0.36\textwidth}
\subfigure[Blogcatalog]
{\includegraphics[width=\textwidth]{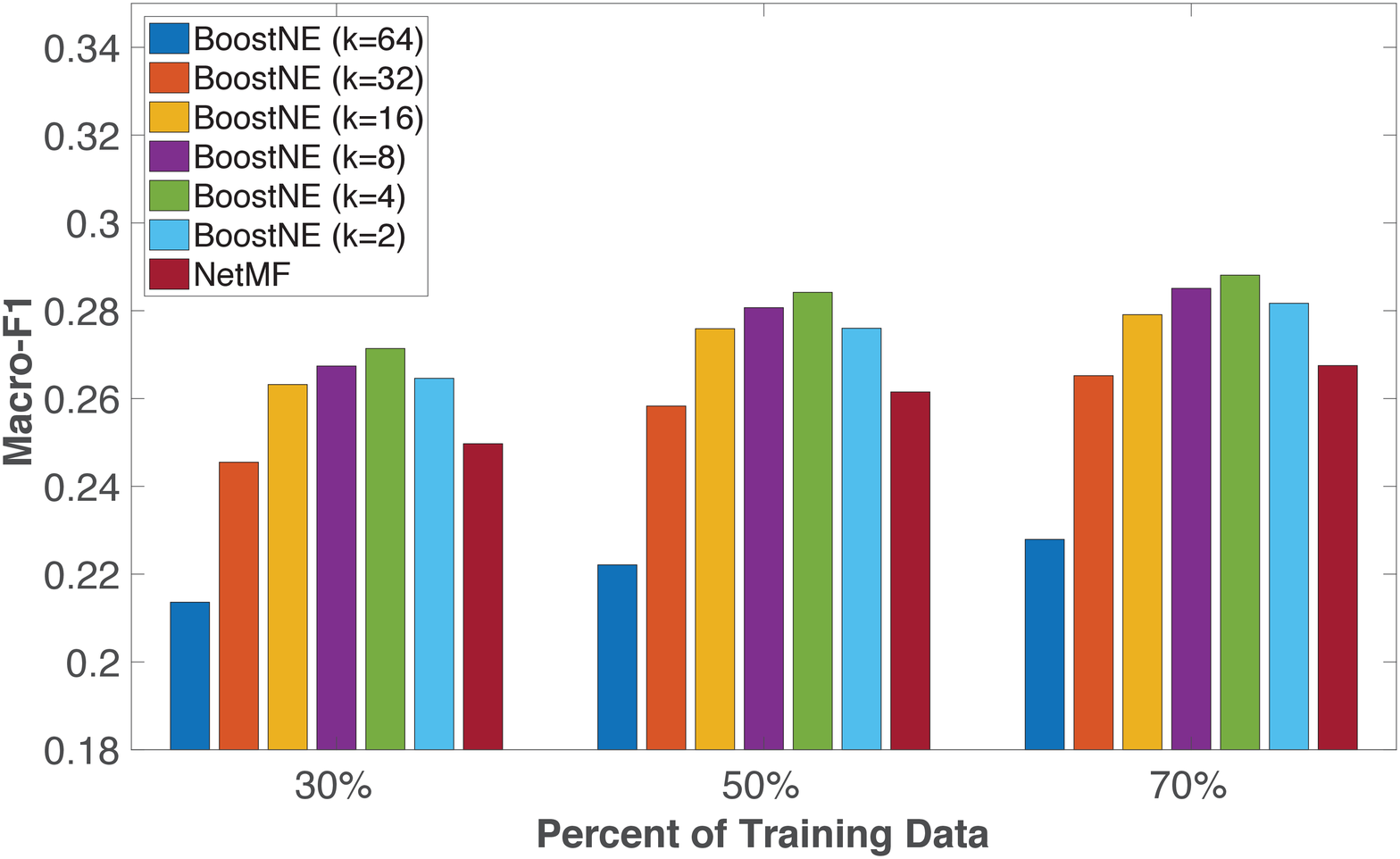}}
\end{minipage}
\centering
\vspace{-0.1in}
\caption{Impact of the number of levels $k$ on the learned node embedding w.r.t. Macro-F1.}
\vspace{-0.2in}
\label{fig:kmacro}
\end{figure*}

\subsection{Further Analysis w.r.t. Approximation Error}
Until now, we have shown that the proposed multi-level network embedding method is superior to NetMF with a single run of NMF on the node connectivity matrix. Despite its empirical success, the underlying reason why performing multiple stages of low-rank approximation leads better embedding representations still remain opaque. In this section, we investigate the underlying mechanism from the approximation error perspective. In particular, we vary the number of levels $k$ in the range of $\{2, 4, 8, 16, 32, 64\}$ and compare the Frobenius norm of the residual matrix after multiple stages of low-rank approximation. The baseline method NetMF can be regarded as a special case when $k=1$. The comparison results on these four datasets are shown in Fig.~\ref{fig:error}.

As can be observed from the figure, the approximation of the node connectivity matrix from BoostNE is much smaller that of NetMF. In particular, the approximation error gradually goes down when the value of $k$ is increased from 2 to 64. This observation helps us explain why the proposed multi-level network embedding framework BoostNE learns better node representations than NetMF. The main reason is that the low-rank assumption of the node connectivity matrix does not always hold in practice and through the multiple stages of low-rank approximation, the node connectivity patterns are better approximated in BoostNE, and further leads more discriminative node embeddings.

\vspace{-0.05in}
\subsection{Impact of the Number of Embedding Levels $k$}
As shown in the previous subsection, larger embedding level $k$ often results in a smaller approximation error of the node connectivity matrix. To further investigate how different values of $k$ impact the learned embeddings, we show the node classification performance w.r.t Micro-F1 and Macro-F1 of BoostNE with different values of $k$ and NetMF in Fig.~\ref{fig:kmicro} and Fig.~\ref{fig:kmacro}. As can be observed from the figures, when we increase the value of $k$, the classification performance first increases then reaches its peak and then gradually decreases. The observations are consistent with different portions of training data (30\%, 50\%, and 70\%) and the best classification performance is achieved when $k$ is set as 4 or 8, where the corresponding $d_{s}$ is 32 or 16. In addition, we also observe that the performance of BoostNE is better than NetMF in a wide range (when $k$ is varied from 2 to 32), which further shows the validity of performing boosted low-rank approximation in learning node embeddings.

\section{Related Work}
We briefly review related work on two aspects: (1) network embedding; and (2) ensemble-based matrix factorization.

\noindent\textbf{Network Embedding} The story of network embedding can be dated back to the early 2000s, when myriad of graph embedding algorithms~\cite{roweis2000nonlinear,belkin2002laplacian,tenenbaum2000global} were developed, as a part of the general dimensionality reduction techniques. Graph embedding first builds an affinity graph based on the feature representations of data instances, and then embeds the affinity graph into a low-dimensional feature space. Even though these algorithms are not explicitly designed for networked data, we can trivially adapt them by feeding the adjacency matrix of networked data as the affinity graph. Along this line, we witnessed a surge of factorization based network embedding methods in recent years, with the target to decompose a carefully designed affinity matrix in capturing the first-order~\cite{belkin2002laplacian,ahmed2013distributed}, higher-order~\cite{cao2015grarep,ou2016asymmetric} node proximity or community structure~\cite{tang2009relational,wang2017community} of the underlying network. Despite their empirical success, the factorization based network embedding methods have at least a quadratic time complexity w.r.t. the number of nodes, prohibiting their practical usage on large-scale networks. The recent advances of network representation learning is largely influenced by the word2vec~\cite{mikolov2013distributed} model in the NLP community. The seminal work of Deepwalk~\cite{perozzi2014deepwalk} first makes an analogy between truncated random walks on a network and sentences in a corpus, and then learns the embedding representations of nodes with the same principle as word2vec. Typical embedding methods along this line include LINE~\cite{tang2015line}, node2vec~\cite{grover2016node2vec} and PTE~\cite{tang2015pte}. A recent work found that the embedding methods with negative sampling (e.g., Deepwalk, LINE, PTE, and node2vec) can be unified into a matrix factorization framework with closed-form solutions~\cite{qiu2018network}, which bridges the gap between these two families of network embedding methods. Aforementioned methods mainly adopt a shallow model and the expressibility of the learned embedding representations are rather limited. As a remedy, researchers also resort to deep learning techniques~\cite{chang2015heterogeneous,wang2016structural,cao2016deep} to learn more complex and nonlinear mapping functions. In addition to the raw network structure, real-world networks are often presented with different properties, thus there is a growing interest to learn the embedding representations of networks from different perspectives, such as attributed network~\cite{yang2015network,huang2017label,huang2017accelerated}, heterogeneous networks~\cite{chen2017task,dong2017metapath2vec}, multi-dimensional networks~\cite{ma2018multi,zhang2018scalable} and dynamic networks~\cite{li2017attributed,zhou2018dynamic}.

\noindent\textbf{Ensemble-based Matrix Factorization} Ensemble methods have shown to be effective in improving the performance of single matrix factorization models, especially in the context of collaborative filtering. DeCoste~\cite{decoste2006collaborative} made one of the first attempts to use ensemble methods to improve the prediction results of a single Maximum Margin Matrix Factorization (MMMF) model. In the sequel, the effectiveness of ensemble models is further validated on the Nystr{\"o}m method~\cite{kumar2009ensemble} and Divide-and-Conquer matrix factorization~\cite{mackey2011divide}. The winner of Netflix Prize~\cite{bell2007modeling,koren2008factorization} advanced the performance of recommendation by capitalizing the advantages of memory-based models and latent factor models. Lee et al.~\cite{lee2012automatic} proposed a feature induction algorithm that works in conjunction with stagewise least-squares, and the combination with induced features is superior to existing methods. Most of the existing efforts on matrix factorization are fundamentally based on the prevalent assumption that the given matrix is low-rank, which is often too restrict in practice. Lee et al. developed a novel framework LLORMA model~\cite{lee2013local,lee2016llorma} to combine the factorization results from multiple locally weighted submatrices, with the assumption that only the local submatrices are low-rank. The success of LLORMA is further extended to other related problems, including social recommendation~\cite{zhao2017collaborative}, topic discovery~\cite{suh2016ensnmf} and image processing~\cite{yao2015colorization}.

\vspace{-0.05in}
\section{Conclusions and Future Work}
Network embedding is a fundamental task in graph mining. Recent research efforts have shown that a vast majority of existing network embedding methods can be unified to the general framework of matrix factorization. Specifically, these methods can be summarized with the following working mechanisms: first, construct the deterministic node connectivity matrix by capturing various types of node interactions, and then apply low-rank approximation techniques to obtain the final node embedding representations. However, the fundamental low-rank assumption of the node connectivity matrix may not necessarily hold in practice, making the resultant low-dimensional node representation inadequate for downstream learning tasks. To address this issue, we relax the global low-rank assumption and propose to learn multiple network representations of different granularity from coarse to fine in a forward stagewise fashion. The superiority of the proposed multi-level network embedding framework over other popular network embedding methods is also in line with the success of gradient boosting
framework, where the ensemble of multiple weak embedding representations (learners) leads to a better and more discriminative one (learner).

Future work can be focused on the following two aspects: first we will provide theoretical results to further understand the principles of the boosted NMF approach; second, we plan to investigate methods that can automatically learn the weights of different levels for more meaningful and discriminative embedding representations.

\balance

\end{document}